# Microwave-Induced Magneto-Intersubband Scattering in a Square Lattice of Antidots

A. A. Bykov[a,b], I. S. Strygin[a,b], A. V. Goran[a], E. E. Rodyakina[a,b], D. V. Nomokonov[a], I. V. Marchishin[a], S. Abedi[c], S. A. Vitkalov[c]

[a] *Rzhanov Institute of Semiconductor Physics, Siberian Branch, Russian Academy of Sciences, Novosibirsk, 630090 Russia*
[b] *Novosibirsk State University, Novosibirsk, 630090 Russia*
[c] *Physics Department, City College of the City University of New York, NY 10031, USA*

The effect of microwave radiation on low-temperature electron magnetotransport in a square antidot lattice with a period of $d \approx 0.8$ μm based on a GaAs quantum well with two occupied energy subbands $E_1$ and $E_2$ is investigated. It is shown that, owing to a significant difference between the electron densities in the subbands, commensurability oscillations of the resistance in the investigated antidot lattice are observed only for the first subband. It is found that microwave irradiation under the cyclotron resonance condition results in the formation of resistance oscillations periodic in the inverse magnetic field in the region of the main commensurability peak. It is established that the period of these oscillations corresponds to the period of magneto-intersubband oscillations. The observed effect is explained by the increase in the rate of intersubband scattering caused by the difference between the electron heating in the subbands $E_1$ and $E_2$.

The effect of microwave radiation on the transport of a two-dimensional (2D) electron gas in antidot lattices based on GaAs quantum wells with one occupied energy subband $E_1$ has been studied since these low dimensional systems were first fabricated [1–3]. The surge of interest in this field of research in the past decade was initiated by the discovery of microwave-induced resistance oscillations in the 2D electron gas at large filling factors [4–6]. The period of these oscillations is determined by the ratio of the circular frequency ω of microwave radiation to the cyclotron frequency $\omega_c = eB/m^*$ (where $B$ is the magnetic field and $m^*$ is the electron effective mass); for this reason, these resistance oscillations are often called $\omega/\omega_c$ oscillations. Recently, $\omega/\omega_c$ oscillations were observed in one-dimensional lateral superlattices [7, 8]. It was shown in these studies that, much as in the case of a "shallow" triangular lattice of antidots [3], geometric commensurability resonances of the resistance and $\omega/\omega_c$ oscillations coexist.

When the second energy subband $E_2$ in the quantum well becomes occupied, the resistance $\rho_{xy}$ is determined by the total concentration $n_H = n_1 + n_2$ of electrons in the subbands and two series of Shubnikov–de Haas (SdH) oscillations will appear in the $\rho_{xx}(B)$ dependence. Transport in a two-subband system is determined both by electron scattering processes within each of the subbands separately and by intersubband scattering [9–13]. The most striking manifestation of intersubband scattering is magneto-intersubband resistance oscillations (MISOs), whose period and amplitude are determined by the expressions

$$E_2 - E_1 = k\hbar\omega_c \qquad (1)$$

and

$$\Delta\rho_{MISO}/\rho_0 = A_{MISO}\lambda^2_{MISO}\cos(2\pi\Delta_{12}/\hbar\omega_c), \qquad (2)$$

respectively. Here, $E_1$ and $E_2$ are the energies of the bottoms of the first and second subbands, respectively; $k$ is a positive integer; $\rho_0 = \rho_{xx}(B=0)$; $A_{MISO} = 2\tau_{tr}/\tau_{12}$, where $\tau_{tr}$ and $\tau_{12}$ are the transport and intersubband scattering times, respectively; $\lambda^2_{MISO} = \lambda_1\lambda_2$, where $\lambda_1 = \exp(-\pi/\omega_c\tau_{q1})$ and $\lambda_2 = \exp(-\pi/\omega_c\tau_{q2})$ are the Dingle factors in the first and second subbands, respectively, where $\tau_{q1}$ and $\tau_{q2}$ are the quantum lifetimes in these subbands; $\lambda_{MISO} = \exp(-\pi/\omega_c\tau_q^{MISO})$, where $\tau_q^{MISO} = 2\tau_{q1}\tau_{q2}/(\tau_{q1} + \tau_{q2})$; and $\Delta_{12} = (E_2 - E_1)$.

To date, $\omega/\omega_c$ oscillations have been observed in both single- and two-band electron systems [14–16]. A distinct feature of the microwave photoresistance of a two-subband electron system, as compared to a single-subband system, is interference between $\omega/\omega_c$ resistance oscillations and MISOs [15]. Here, we present the first results of experimental studies of the microwave photoresistance of the two-subband electron system in a square antidot lattice based on a single GaAs quantum well with two occupied energy subbands.

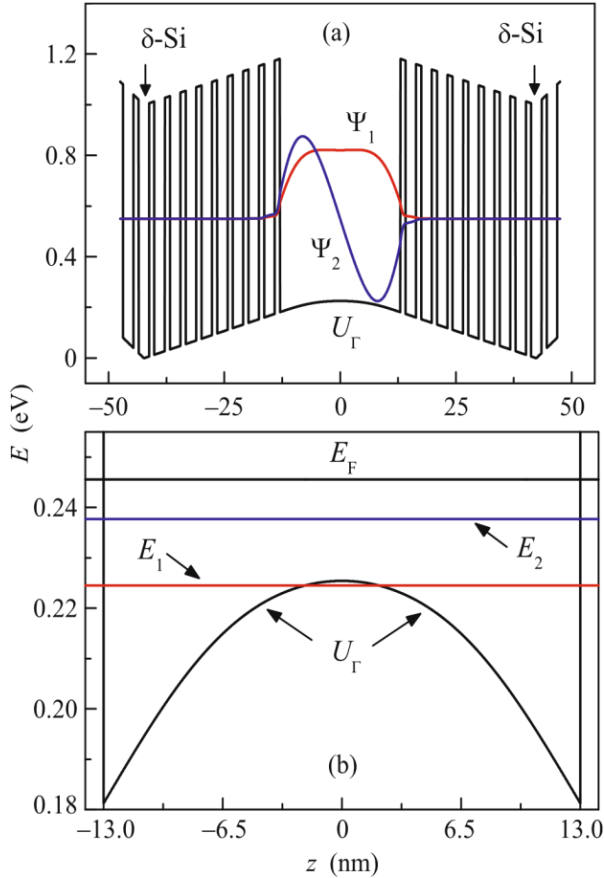

**Fig. 1.** (a) Potential profile of the bottom of the conduction band at the Γ point and the wavefunctions $\Psi_1(z)$ and $\Psi_2(z)$ of the symmetric and antisymmetric energy states, respectively, in a GaAs quantum well confined by AlAs/GaAs superlattice barriers. Arrows indicate the locations of δ-doped layers in the barriers. (b) z-Coordinate dependences of $U_\Gamma$, $E_1$, $E_2$, and $E_F$.

The main goal of this study was to establish the role of intersubband scattering in the microwave photoresistance of a quasi-two-dimensional electron gas in the square lattice of antidots in classically strong magnetic fields.

The original heterostructure was a single 26-nm wide GaAs quantum well confined by AlAs/GaAs superlattice barriers [17–19]. The results of self-consistent calculations of electron wavefunctions for the first two energy subbands in the GaAs quantum well and the band diagram of the heterostructure are presented in Fig. 1. The energy splitting $\Delta_{12} = E_2 - E_1$ obtained from these calculations is about 13.2 meV.

The GaAs quantum well was populated with charge carriers by δ-Si doping. Single δ-Si layers were introduced on both sides of the quantum well at a distance of 29.4 nm from its edges. The distance from the center of the quantum well to the planar surface of the structure was 117.7 nm. The heterostructure was grown by molecular beam epitaxy on a GaAs (100) substrate.

The measurements were carried out on Hall bars with a width of $W = 50$ μm and a length of $L = 100$ μm fabricated by optical photolithography and wet etching. A simplified layout of the sample is shown in the inset of Fig. 2a. The sample consists of two Hall bars, with a square lattice of antidots formed on one of them. The lattice period and the diameter of the antidots were $d \approx 800$ nm and $a \approx 200$ nm, respectively. The lattice was fabricated by electron-beam lithography and dry etching. The experiments were carried out at a temperature of $T = 4.2$ K in magnetic fields $B < 2$ T. The resistance of the samples was measured using an alternating current with a frequency of 762 Hz whose magnitude did not exceed $10^{-6}$ A. The electron density and mobility in the original heterostructure were $n_H \approx 8.1 \times 10^{15}$ m$^{-2}$ and μ $\approx 73$ m$^2$/(V s), respectively. Microwave radiation was delivered to the sample through a circular waveguide with a diameter of 6 mm. The sample was located at a distance of a few millimeters from the open end of the waveguide.

Figure 2a shows the dependence of the resistance $R_{23}$ on the magnetic field $B$ measured for the reference bar at a temperature of $T = 4.2$ K. Magneto-intersubband oscillations are observed in this dependence in the range of magnetic fields 0.1 T $< B <$ 0.6 T, and in magnetic fields of $B > 0.6$ T, MISOs coexist with SdH oscillations [17]. The Fourier spectrum of this $R_{23}(1/B)$ dependence features three frequencies. Two of them correspond to the frequencies of SdH oscillations in the first and second subbands ($f_{SdH1} \approx 12.9$ T and $f_{SdH2} \approx 4.0$ T), and the third corresponds to MISOs ($f_{MISO} \approx 8.9$ T). The electron densities in the subbands calculated from the SdH oscillation frequencies are $n_1 \approx 6.2 \times 10^{15}$ m$^{-2}$ and $n_2 \approx 1.9 \times 10^{15}$ m$^{-2}$. The intersubband splitting determined from the frequency $f_{MISO}$ is $\Delta_{12} \approx 15.2$ meV, which is close to the value obtained from self-consistent calculations for the quantum well under study.

Figure 2b shows the dependence of the resistance $R_{45}$ on the magnetic field $B$ measured for the bar with a square lattice of antidots at a temperature of $T = 4.2$ K. This dependence features two prominent peaks marked with arrows at $B_1 \approx 0.31$ T and $B_2 \approx 0.11$ T. Similar peaks were for the first time observed in the magnetoresistance of single-subband electron systems in square lattices of antidots and were explained by the commensurability of the electron cyclotron radius $R_c$ and the lattice period $d$ [20–23].

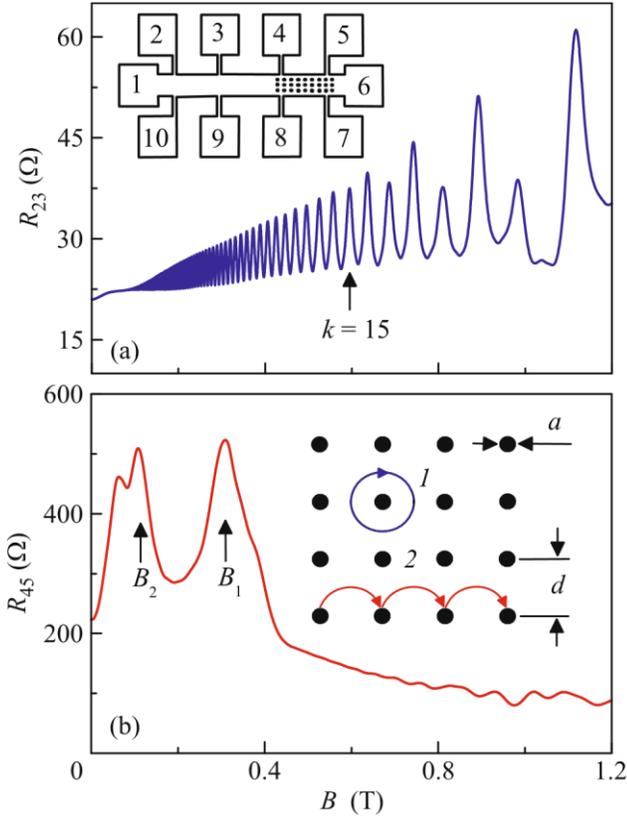

**Fig. 2.** (a) Magnetic-field dependence $R_{23}(B)$ recorded for the reference bar at a temperature of $T = 4.2$ K. The arrow shows the position of the magnetointersubband peak number $k = 15$. The inset shows the layout of the sample with one section patterned with an antidot lattice. (b) Magnetic-field dependence $R_{45}(B)$ recorded for the bar with a square lattice of antidots at a temperature of $T = 4.2$ K. Arrows indicate the magnetic fields $B_1$ and $B_2$ corresponding to $R_{c1}/d \approx 0.5$ and 1.5, respectively. The inset shows two characteristic types of electron trajectories in a square antidot lattice: (*1*) pinned orbit and (*2*) runaway trajectory.

Geometric resonances in the resistance occur because electrons moving along pinned and runaway stable trajectories make a significant contribution to the classical magnetotransport in antidot lattices [22, 23]. In the square lattice of antidots, the two most prominent commensurability peaks of the resistance occur in magnetic fields corresponding to $R_c/d \approx 0.5$ and 1.5.

In much the same way as two-subband magnetotransport in one-dimensional lateral superlattices [24, 25], two series of geometric resonances for each of the subbands should be observed in the $R_{45}(B)$ dependence. However, this dependence features only two commensurability peaks, as in the case of a single-subband electron system in a square antidot lattice [21]. This can be explained by the fact that $n_1$ is much larger than $n_2$ in the antidot lattice under study. In this case, the contribution of the second subband to magnetotransport will be much smaller than that of the first subband. The electron densities in the subbands for the bar with an antidot lattice calculated from the frequencies of the SdH oscillations are $n_1 \approx 5.9 \times 10^{15}$ m$^{-2}$ and $n_2 \approx 1.8 \times 10^{15}$ m$^{-2}$; i.e., $n_1 > 3n_2$, which is consistent with our interpretation. The magnetic fields corresponding to the conditions $R_{c1}/d \approx 0.5$ and $R_{c1}/d \approx 1.5$ are $B_1 \approx 0.32$ T and $B_2 \approx 0.11$ T, respectively. These calculated values coincide well with the experimental ones, which indicates that the contribution of the first subband to classical magnetotransport in the antidot lattice under study is dominant.

Figure 3a shows the effect of microwave radiation on the resistance of the two-subband electron system. In contrast to single-subband systems, we observe interference between $\omega/\omega_c$ oscillations and MISOs. This behavior is explained by the influence of intersubband coupling on the frequency-dependent photoinduced part of the nonequilibrium electron distribution function [15]. Figure 3b shows a different effect of microwave radiation on the resistance of the twosubband electron system in a square antidot lattice. There are no $\omega/\omega_c$ oscillations in the $R_{45}^{\omega}(B)$ dependence, and the amplitude of commensurability peaks is suppressed significantly. We attribute the suppression of commensurability oscillations to the heating of the electron gas by microwave radiation and the absence of $\omega/\omega_c$ oscillations to the scattering of electrons by the antidot potential. It is unusual that an oscillating component, which is absent in the $R_{45}(B)$ dependence, appears in the $R_{45}^{\omega}(B)$ dependence under the conditions of the cyclotron resonance.

The behavior of the $R_{45}(B)$ and $R_{45}^{\omega}(B)$ dependences in the region of the main commensurability peak is shown in detail in Fig. 4a. It can be seen that the $R_{45}^{\omega}(B)$ dependence, in contrast to the $R_{45}(B)$ dependence, exhibits an oscillating component. The analysis of this component demonstrates that oscillations are periodic in $1/B$ and their period corresponds to MISOs.

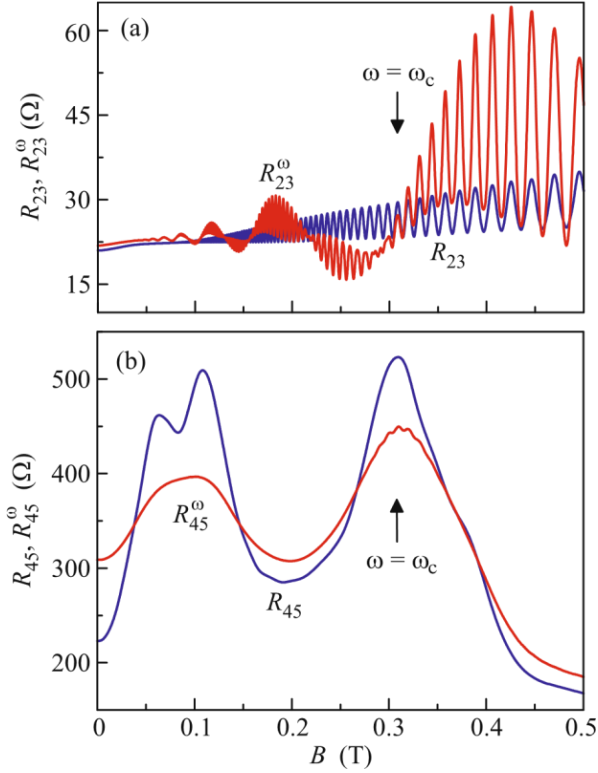

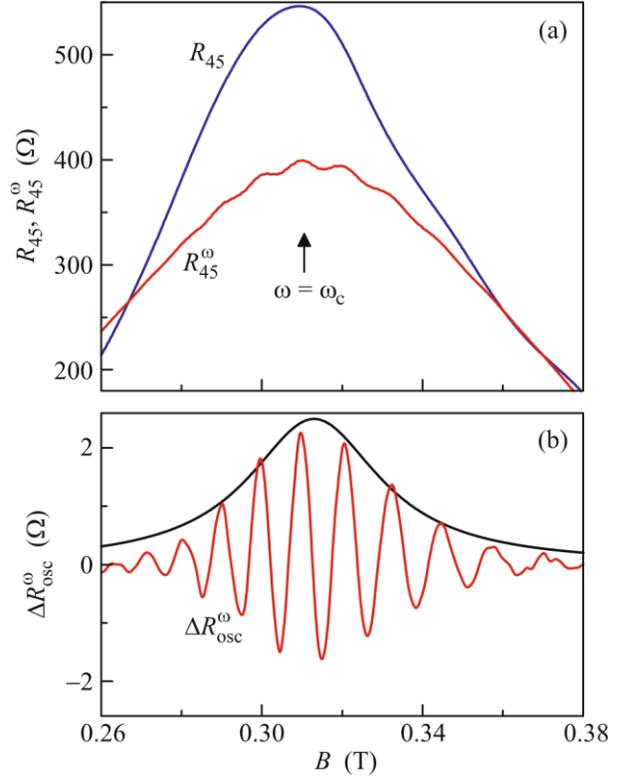

**Fig. 3.** (a) Magnetic-field dependences $R_{23}(B)$ and $R_{23}^{\omega}(B)$ recorded for the reference bar at a temperature of $T = 4.2$ K without microwave irradiation and under microwave irradiation at a frequency of $\omega/2\pi \approx 150$ GHz, respectively. The arrow shows the position of the cyclotron resonance. (b) Magnetic-field dependences $R_{23}(B)$ and $R_{23}^{\omega}(B)$ recorded for the bar with a square lattice of antidots at a temperature of $T = 4.2$ K without microwave irradiation and under microwave irradiation at a frequency of $\omega/2\pi \approx 150$ GHz, respectively. The arrow shows the position of the cyclotron resonance.

**Fig. 4.** (a) Magnetic-field dependences $R_{45}(B)$ and $R_{45}^{\omega}(B)$ near the main commensurability peak recorded for the bar with a square lattice of antidots at a temperature of $T = 4.2$ K without microwave irradiation and under microwave irradiation at a frequency of $\omega/2\pi \approx 150$ GHz, respectively. (b) (Thick red line) Oscillating component $\Delta R_{\mathrm{osc}}^{\omega}(B)$ of the resistance and (thin black line) Lorentzian curve calculated with a fitting parameter $\tau_{\mathrm{CR}} = 9.7$ ps.

It is surprising that MISOs occur under microwave irradiation. According to Eq. (2), the amplitude of MISOs depends exponentially on $1/\tau_q^{\mathrm{MISO}}$. Scattering by antidots reduces $\tau_q^{\mathrm{MISO}}$, which results in the suppression of MISOs. Heating of the electron gas should lead to the same result, because an increase in temperature reduces $\tau_q^{\mathrm{MISO}}$ [17]. However, we observe experimentally that, on the contrary, microwave irradiation under the cyclotron resonance conditions restores MISOs.

In our opinion, the appearance of MISOs in the antidot lattice in the region of the main commensurability peak under microwave irradiation is caused by the enhancement of intersubband scattering, which takes place because electron heating in subbands $E_1$ and $E_2$ leads to different nonequilibrium temperatures $T_{\mathrm{e}1}$ and $T_{\mathrm{e}2}$ in these subbands.

In the magnetic field $B_1$, a significant fraction of electrons in subband $E_1$ move along pinned orbits and avoid collisions with antidots, in contrast to electrons in subband $E_2$. Therefore, the cyclotron resonance line for electrons in the former subband will be narrower than that for electrons in the latter subband. Thus, the absorption of microwave radiation under the conditions of cyclotron resonance near the magnetic field $B_1$ will be more efficient in the first than in the second subband, and, accordingly, $T_{\mathrm{e}1}$ will be higher than $T_{\mathrm{e}2}$.

We believe that the rate of elastic electron transitions between subbands when $T_{\mathrm{e}1} = T_{\mathrm{e}2}$ will change insignificantly compared to the equilibrium situation. In this case, the microwave heating of the two-subband electron system in an antidot lattice should only reduce $\tau_q^{\mathrm{MISO}}$ owing to the enhancement of electron–electron scattering. However, when $T_{\mathrm{e}1} > T_{\mathrm{e}2}$, the rate of elastic electron transitions between the subbands at the edges of the energy interval $k_{\mathrm{B}} T_{\mathrm{e}1}$ should increase, which will reduce $\tau_{12}$.

In the energy range between $E_F + k_B T_{e1}/2$ and $E_F + k_B T_{e2}/2$, the rate of transitions from the first to the second subband will increase, since the energy states in this range in the first subband are occupied and those in the second subband are free. In the energy range between $E_F - k_B T_{e1}/2$ and $E_F - k_B T_{e2}/2$, additional electron transitions occur from the second to the first subband, since there are free energy states for such transitions in the first subband.

Figure 4b shows that, as with the shape of a classical resonance line, the magnetic-field dependence of the amplitude of the observed MISOs is well described by a Lorentzian curve with one fitting parameter $\tau_{CR}$. The found value $\tau_{CR} = 9.7$ ps is several times longer than the transport time of electron scattering in the antidot lattice $\tau_{tr}^{SAL} \approx 2.9$ ps. This relation between $\tau_{tr}^{SAL}$ and $\tau_{CR}$ indicates that the width of the resonance line is determined by electrons in the first subband, which do not collide with antidots. Magneto-intersubband oscillations have the maximum amplitude under the conditions of resonance absorption of microwaves, when $T_{e1}$ is maximal. Under these conditions, the difference $T_{e1} - T_{e2}$ also has the maximum value, and, correspondingly, the rate of intersubband transitions is the highest.

In summary, we have investigated magnetotransport in a two-subband electron system in a square antidot lattice. We have demonstrated that commensurability oscillations of the resistance in the lattice under study are observed only for the first subband, where the electron density is higher than that in the second subband. We have established that the amplitude of commensurability oscillations in the antidot lattice under microwave irradiation decreases and $\omega/\omega_c$ oscillations do not occur. We found that, under the cyclotron resonance conditions, resistance oscillations caused by magneto-intersubband scattering appear in the region of the main commensurability peak. The magnetic-field dependence of the amplitude of the observed magneto-intersubband oscillations is quantitatively described by a Lorentzian curve whose width is determined by a relaxation time that is longer than the transport time of electron scattering in the antidot lattice. The observed effect is explained by a decrease in the intersubband scattering time when the electron subsystems in the two subbands are heated by microwave radiation with different efficiencies.

FUNDING

This study was supported by the Russian Foundation for Basic Research (project no. 18-02-00603) and the Division of Materials Research, US National Science Foundation (grant no. 1702594).

REFERENCES

[1] A. A. Bykov, G. M. Gusev, Z. D. Kvon, V. M. Kudryashev, and V. G. Plyukhin, JETP Lett. **53**, 427 (1991).
[2] E. Vasiliadou, R. Fleischmann, D. Weiss, D. Heitmann, K. V. Klitzing, T. Geisel, R. Bergmann, H. Schweizer, and C. T. Foxon, Phys. Rev. B **52**, R8658(R) (1995).
[3] Z. Q. Yuan, C. L. Yang, R. R. Du, L. N. Pfeiffer, and K. W. West, Phys. Rev. B **74**, 075313 (2006).
[4] M. A. Zudov, R. R. Du, J. A. Simmons, and J. L. Reno, Phys. Rev. B **64**, 201311(R) (2001).
[5] P. D. Ye, L. W. Engel, D. C. Tsui, J. A. Simmons, J. R. Wendt, G. A. Vawter, and J. L. Reno, Appl. Phys. Lett. **79**, 2193 (2001).
[6] I. A. Dmitriev, A. D. Mirlin, D. G. Polyakov, and M. A. Zudov, Rev. Mod. Phys. **84**, 1709 (2012).
[7] A. A. Bykov, I. S. Strygin, E. E. Rodyakina, W. Mayer, and S. A. Vitkalov, JETP Lett. **101**, 703 (2015).
[8] A. A. Bykov, I. S. Strygin, A. V. Goran, A. K. Kalagin, E. E. Rodyakina, and A. V. Latyshev, Appl. Phys. Lett. **108**, 012103 (2016).
[9] V. M. Polyanovskii, Sov. Phys. Semicond. **22**, 1408 (1988).
[10] P. T. Coleridge, Semicond. Sci. Technol. **5**, 961 (1990).
[11] M. E. Raikh and T. V. Shahbazyan, Phys. Rev. B **49**, 5531 (1994).
[12] N. S. Averkiev, L. E. Golub, S. A. Tarasenko, and M. Willander, J. Phys.: Condens. Matter **13**, 2517 (2001).
[13] O. E. Raichev, Phys. Rev. B **78**, 125304 (2008).
[14] A. A. Bykov, D. R. Islamov, A. V. Goran, and A. I. Toropov, JETP Lett. **87**, 477 (2008).
[15] S. Wiedmann, G. M. Gusev, O. E. Raichev, T. E. Lamas, A. K. Bakarov, and J. C. Portal, Phys. Rev. B **78**, 121301(R) (2008).
[16] A. A. Bykov, A. V. Goran, and A. K. Bakarov, J. Phys.D: Appl. Phys. **51**, 28LT01 (2018).
[17] A. V. Goran, A. A. Bykov, A. I. Toropov, and S. A. Vitkalov, Phys. Rev. B **80**, 193305 (2009).
[18] A. A. Bykov, A. V. Goran, and S. A. Vitkalov, Phys.Rev. B **81**, 155322 (2010).
[19] A. A. Bykov, JETP Lett. **100**, 786 (2015).
[20] K. Ensslin and P. M. Petroff, Phys. Rev. B **41**, 12307(R) (1990).
[21] D. Weiss, M. L. Roukes, A. Menschig, P. Grambow, K. von Klitzing, and G. Weimann, Phys. Rev. Lett. **66**, 2790 (1991).
[22] R. Fleischmann, T. Geisel, and R. Ketzmerick, Phys.Rev. Lett. **68**, 1367 (1992).
[23] E. M. Baskin, G. M. Gusev, Z. D. Kvon, A. G. Pogosov, and M. V. Entin, JETP Lett. **55**, 678 (1992).
[24] J. P. Lu and M. Shayegan, Phys. Rev. B **58**, 1138 (1998).
[25] A. A. Bykov, I. S. Strygin, A. V. Goran, D. V. Nomokonov, I. V. Marchishin, A. K. Bakarov, E. E. Rodyakina, and A. V. Latyshev, JETP Lett. **110**, 354 (2019).